\documentstyle[12pt,epsfig,aps,prd,eqsecnum]{revtex}

\begin{document}
 
\renewcommand{\theequation}{\arabic{section}.\arabic{equation}}

\title{General Green's function formalism for transport calculations with 
$spd$-Hamiltonians and giant magnetoresistance
in Co and Ni based magnetic multilayers
}
\author{S. Sanvito\thanks{e-mail: sanvito@dera.gov.uk},
C.J. Lambert\thanks{e-mail:c.lambert@lancaster.ac.uk},}
\address{School of Physics and Chemistry, Lancaster University LA1 4YB,
Lancaster UK}
\author{J.H. Jefferson,}
\address{DERA, Electronic and Optical Materials Centre, Malvern,
Worcs. WR14 3PS, UK}
\author{A.M. Bratkovsky\thanks{e-mail: alexb@hpl.hp.com}}
\address{Hewlett-Packard Laboratories, 3500 Deer Creek Road,
Palo Alto, CA 94304-1392}
\date{\today}
\maketitle
\begin{abstract}
A novel, general Green's function
technique for elastic spin-dependent transport calculations is
presented, which (i) scales linearly with system size and 
(ii) allows straightforward application to
general tight-binding Hamiltonians ($spd$ in the present work). 
The method is applied to studies of  conductance and 
giant magnetoresistance 
(GMR) of magnetic multilayers in CPP (current perpendicular to planes)
geometry in the limit of large coherence length.
The magnetic materials considered are Co and Ni, 
with various non-magnetic materials from the
3$d$, 4$d$, and 5$d$ transition metal series. Realistic tight-binding
models for them have been constructed with the use of density functional
calculations. 
We have identified three qualitatively different cases which depend on
whether or not  the bands (densities of states) of a non-magnetic metal
        (i) form an almost perfect match with one of spin sub-bands
of the magnetic metal (as in Cu/Co spin valves); 
        (ii) have almost pure  $sp$ character at the Fermi level
(e.g. Ag); 
        (iii) have almost pure $d$ character at the Fermi energy (e.g. Pd, Pt).
The key parameters which give rise to a large GMR ratio turn out to be 
        (i) a strong spin polarization of the magnetic metal, 
        (ii) a large energy offset between the conduction band 
of the non-magnetic metal and one of spin sub-bands of the 
magnetic metal, and 
        (iii)  strong interband scattering in one of spin
sub-bands of a magnetic metal. 
The present results show that GMR oscillates with variation of 
the thickness of either non-magnetic or magnetic layers, 
as observed experimentally.

\end{abstract}


\section{INTRODUCTION}

The discovery \cite{bab} of giant magnetoresistance 
(GMR) in metallic multilayers about a decade ago
has attracted a great deal of attention. This is not only because of the
possibility of building
sensitive magnetometers, but also because
GMR provides valuable insight into
spin-dependent transport in inhomogeneous systems. GMR is the drastic change
in  electrical resistance that occurs when a strong magnetic field is applied
to a superlattice made with alternating
 magnetic and non-magnetic (spacer) metallic layers.
Early experiments were conducted with
the so-called current-in-plane (CIP) configuration, in which the current flows
parallel to the  plane of the layers. In this configuration the dimensions
 of the system are macroscopic
with transport properties being reasonably described by a classical Boltzmann equation
\cite{barnas}
and GMR is associated with the spin-dependent scattering of electrons 
at the interfaces.

The first experiments with the current perpendicular to the plane of the
layers (CPP) \cite{prat} paved the way  to a completely different
regime, where quantum effects can play a dominant role.
 In good quality superlattices
the elastic mean free path can extend over several layers
 and the spin diffusion
length can be longer than the total superlattice thickness. 
In this case we can talk, according to Mott \cite{mott},  about majority and
minority spin carriers as two independent spin fluids which remain  coherent as they
cross the superlattice layers.  Then a full quantum description is required.
For such structures
the magnetoconductance  and GMR are found to be oscillatory functions
of the non-magnetic layer thickness \cite{parkin2} with periods extending over
several atomic planes. 
Despite the evidence of such important quantum effects,
early theoretical work was based on spin-dependent scattering 
at interfaces and/or
magnetic impurities and completely neglected quantum interference.
In 1995 Schep et al. \cite{shep,shep2} challenged this conventional picture
and showed that
large values of GMR (of order 120\%)
exist even in absence of impurity scattering.

The aim of the present paper is to develop a quantitative approach to
quantum transport, which describes the dependence of
GMR on specific material and/or layer thickness.
Our calculations are based on the Landauer-B\"uttiker formalism \cite{but},
using a nearest neighbor tight-binding $spd$ Hamiltonian.
The tight-binding energy parameters
have been fitted to accurate {\it ab-initio} density functional 
calculations.

We have developed a novel and completely general technique for calculating
Green functions and hence a scattering $S$ matrix and transport
coefficients
of a  finite
superlattice connected to pure crystalline semi-infinite leads.
This allowed us to perform a systematic study of
Co/A and Ni/A 
multilayers, where A=Cu, Ag, Pd, Au and Pt,
and analyze  
optimal conditions for GMR.

We shall consider below a
GMR ratio of a finite superlattice connected to two semi-infinite crystalline
leads \cite{bauer}, as sketched in Fig.~1.
The GMR ratio is defined by
\begin{equation}
\rm{GMR}=(\Gamma_{\rm{FM}}^{\uparrow}+
\Gamma_{\rm{FM}}^{\downarrow}
-2\Gamma_{\rm{AF}}^{\uparrow\downarrow})/
2\Gamma_{\rm{AF}}^{\uparrow\downarrow}\;{,}
\label{gmr}
\end{equation}
where $\Gamma_{\rm{FM}}^{\sigma}$ is the conductance of a given
spin channel
$\sigma$ in the ferromagnetic (FM)  configuration and
$\Gamma_{\rm{AF}}^{\uparrow\downarrow}$ is the corresponding conductance
(for either spin)
in the anti-ferromagnetic state.
We calculate the conductance by evaluating the Landauer
formula \cite{but} 

\begin{equation}
\Gamma^{\sigma}=\frac{e^2}{h}T^\sigma\;{,}
\label{butlan}
\end{equation}
where $T^\sigma$ is  the total transmission coefficient for the
spin $\sigma$ calculated at the Fermi energy. In what follows
we assume a perfect match at the interface between the fcc lattices
of the different metals. This assumption is
particularly good in the case of Co, Cu, and Ni which possess
almost identical lattice constants. Equation (\ref{butlan})
is valid even in the presence of disorder. 
We shall consider below crystalline systems with smooth interfaces, where
 $k_\parallel$ is a good quantum number (we use the
symbol $\parallel$ for the in-plane coordinates and $\perp$
for the direction of the current).  The Hamiltonian can then 
be diagonalized
in the Bloch basis $k_{\parallel}$ to yield

\begin{equation}
\Gamma^{\sigma}=\sum_{k_{\parallel}}\Gamma^{\sigma}(k_{\parallel})=
\frac{e^2}{h}\sum_{k_{\parallel}}T^\sigma(k_{\parallel})\;{,}
\label{blocon}
\end{equation}
where the sum over $k_{\parallel}$ is extended over the two-dimensional 
Brillouin zone in the case of infinite cross section and over the allowed  
discrete $k_{\parallel}$'s in the case of finite cross section.

\section{
		Green's function formalism for
		scattering matrix.
}

The problem of analyzing a realistic tight-binding model, with $spd$ orbitals
on each site represents a formidable numerical challenge \cite{oxo,matto}.
In this section we describe a very general and efficient
technique for calculating
the $S$-matrix and hence the transmission coefficients of an arbitrary
scattering region such
as a superlattice,  attached to
two semi-infinite crystalline leads. The key components of the calculation
are i) the retarded Green function $g$ of the semi-infinite leads and ii) an effective
Hamiltonian $H_{\rm{eff}}$ describing the scattering region
and its coupling to the leads.
The approach described below
provides a versatile method for computing these two components,
which are then combined via Dyson's equation to yield the Green function of
the complete structure.
A novel feature of the technique is that it
avoids adding a small imaginary part to the energy and
provides a semi-analytic formula for $g$.

\subsection{The Green function of an arbitrary semi-infinite lead.}

To compute the Green function for a
semi-infinite crystalline lead of finite cross-section,
we first calculate the Green function of a doubly
infinite system and then  derive the semi-infinite case by applying
boundary conditions at the end of the lead.
To this end, consider the doubly infinite system shown in Fig.\ref{greensch}.

If $z$ is the direction of transport,
the system comprises a periodic sequence of slices, described by an
intra-slice matrix $H_0$ and coupled by
a nearest neighbor inter-slice hopping matrix $H_1$.  The nature of
the slices need not be specified at this stage. They can describe a
single band atom in an atomic chain, an
atomic plane or a more complex cell.
For such a general system, the total 
Hamiltonian $H$ can be written as an 
infinite matrix of the form

\begin{equation}
H=\left(
\begin{array}[h]{rrrrrrrr}
... & ... & ... & ... & ... & ... & ... & ... \\
... & H_0 & H_1 & 0 & ... & ...  & ... & ... \\
... & H_{-1} & H_0 & H_1 & 0 & ... & ... & ... \\
... & 0 & H_{-1} & H_0 & H_1 & 0 & ... & ... \\
... & 0 & 0 & H_{-1} & H_0 & H_1 & 0  & ... \\
... & ... & ... & ... & ... & ... & ... & ... \\
... & ... & ... & ... & ... & ... & ... & ... \\
\end{array}
\right){\;},
\label{ham}
\end{equation}

where $H_0$ is Hermitian and
$H_{-1}=H_1^\dagger$.
The Schr\"odinger equation for this system is of the form

\begin{equation}
H_0\psi_z+H_1\psi_{z+1}+H_{-1}\psi_{z-1}=E\psi_z{\;},
\label{sch}
\end{equation}

where $\psi_z$ is a column vector corresponding to the slice at 
the position $z$ with $z$ an integer measured in units of inter-slice
distance. Let the quantum numbers corresponding to the degrees of freedom
within a slice be $\mu=1,2,\ldots,M$ and the corresponding
components of $\psi_z$ be $\psi_z^\mu$. For example in the following
sections, these enumerate the atomic sites
within the slice and the valence orbitals ($spd$) at a site. 
The Schr\"odinger equation may then be solved by 
introducing the Bloch state, 

\begin{equation}
\psi_z=n_{k_\perp}^{1/2}e^{ik_\perp{z}}\phi_{k_\perp}{\;},
\label{jwf}
\end{equation}
where $\phi_{k_\perp}$ is a normalized M-component column
vector and $n_{k_\perp}^{1/2}$ an arbitrary constant.
 Substituting into the
equation (\ref{sch}) gives
\begin{equation}
\left(H_0+H_1e^{ik_\perp}+H_{-1}e^{-ik_\perp}-E\right)
{\phi}_{k_\perp}=0{\;},
\label{scsum2}
\end{equation}

Our task is to compute the Green function $g$ of such a structure, for
all real energies.
For a given energy $E$, the first task is to
determine all possible values (both real and complex)
of the wavevectors $k_{\perp}$
by solving the secular equation

\begin{equation}
\det(H_0+H_1\chi+H_{-1}/\chi-E)=0{\;}.
\label{dislaw}
\end{equation}
where $\chi=e^{ik_\perp}$.

In contrast to conventional band-theory, where
the problem is to compute the $M$ values of $E$ for a given (real) choice
of $k_\perp$, our aim is to compute the complex roots $\chi$
of the polynomial (\ref{dislaw}) for a given (real) choice of $E$.
Consider first the case where $H_1$ is not singular.  
We note that for real $k_\perp$,
conventional band-theory yields $M$ energy bands $E_n(k_\perp)$,
$n=1,\ldots,M$, with $E_n(k_\perp+2\pi)=E_n(k_\perp)$.
As a consequence, for a given
choice of $E$, to each real solution $k_\perp = k$, for which
the group velocity
\begin{equation}
{v_k}=\frac{1}{\hbar}
\frac{\partial{E}(k)}{\partial{k}}
{\;}
\label{gvbcc}
\end{equation}
is positive,
there exists a second solution
$k_\perp = \bar{k}$ for which the group
velocity
\begin{equation}
{v_{\bar k}}=\frac{1}{\hbar}
\frac{\partial{E}(\bar{k})}{\partial{\bar{k}}}
{\;}
\label{gvbcc1}
\end{equation}
is negative. These real wavevectors define the open
scattering channels of the structure and
in the simplest case, where $H_1 = H_{-1}$, one finds $k=-\bar k$.
We also note that to each solution $k_\perp$ the Hermitian conjugate of
(\ref{scsum2}) shows that $k_\perp^*$ is also a solution.
Hence to each \lq\lq right-decaying" solution $k$ possessing a positive
imaginary part, there is a \lq\lq left-decaying" solution $\bar k$
with a negative imaginary part.
For the purpose of constructing the Green function,
this leads us to
 divide the roots of (\ref{scsum2}) into two sets: the first set
of $M$ wavevectors labeled $k_l$ ($l= 1,...,M$) correspond to right-moving
and right-decaying plane-waves and the second set
labeled $\bar k_l$ ($l= 1,...,M$) correspond to left-moving
and left-decaying plane-waves.

Although the solutions to (\ref{dislaw}) can be found using a root tracking
algorithm, for numerical purposes it is more convenient to map
(\ref{scsum2}) onto an equivalent eigenvalue problem by introducing the matrix
$\cal{H}$ 

\begin{equation}
\cal{H}=\left(
\begin{array}[4]{rr}
-H_1^{-1}({H_0}-E) & -H_1^{-1}H_{-1} \\
{\cal I}\;\;\;\;\;\;\;\;\; & 0\;\;\;\;\;\;\;\;\; \\
\end{array}
\right){\;},
\label{hacca}
\end{equation}
where ${\cal I}$ is the $M$ dimensional identity matrix.
The eigenvalues of $\cal{H}$ are the $2M$ roots
$e^{ik_l}$, $e^{i\bar k_l}$  and the upper $M$ components
of the eigenvectors of $\cal{H}$ are the  corresponding
eigenvectors ${\phi}_{k_l}, {\phi}_{\bar{k}_l}$.

To construct the retarded Green function $g_{zz^\prime}$ of
the doubly infinite system, we note that except at $z=z^\prime$, $g$ is simply
a wavefunction and hence must have the form

\begin{equation}
g_{zz^\prime}=\left\{
\begin{array}[4]{r}
\sum_{l=1}^M{\phi}_{k_l}e^{ik_l(z-z^\prime)}
{\mathsf w}_{k_l}^\dag\;\;\;\;\;z\geq{z}^\prime \\
\\
\sum_{l=1}^M{\phi}_{\bar{k}_l}e^{i\bar{k}_l(z-z^\prime)}
{\mathsf w}_{\bar{k}_l}^\dag
\;\;\;\;\;z\leq{z}^\prime \\
\end{array}
\right.\;
\label{gf}
\end{equation}

where the $M$-component vectors ${\mathsf w}_{k_l}$
and ${\mathsf w}_{\bar{k}_l}$ are  to be determined.
Since $g_{zz^\prime}$ is  retarded
both in $z$ and $z^\prime$,
it satisfies the Green function equation corresponding to  (\ref{sch})
and is
continuous at the point $z=z^\prime$, one obtains 

\begin{equation}
g_{zz^\prime}=\left\{
\begin{array}[4]{r}
\sum_{l=1}^M{\phi}_{k_l}e^{ik_l(z-z^\prime)}
\tilde{{\phi}}_{k_l}^\dag{\cal{V}}^{-1}\;\;\;\;\;z\geq{z}^\prime \\
\\
\sum_{l=1}^M{\phi}_{\bar{k}_l}e^{i\bar{k}_l(z-z^\prime)}
\tilde{{\phi}}_{\bar{k}_l}^\dag{\cal{V}}^{-1}
\;\;\;\;\;z\leq{z}^\prime \\
\end{array}
\right.\;{.}
\label{digf}
\end{equation}

The matrix ${\cal{V}}$ is defined by

\begin{equation}
{\cal{V}}=\sum^{M}_{l=1}H_{-1}\left[
{\phi}_{k_l}e^{-ik_l}\tilde{{\phi}}_{k_l}^\dag-
{\phi}_{\bar{k}_l}e^{-i\bar{k}_l}\tilde{{\phi}}_{\bar{k}_l}^\dag
\right]\;{,}
\label{v}
\end{equation}

and the set of vectors $\tilde{{\phi}}_{k_l}^\dag$ 
($\tilde{{\phi}}_{\bar{k}_l}^\dag$) are the duals of the set
${\phi}_{k_l}$ (${\phi}_{\bar{k}_l}$), defined by

\begin{equation}
\tilde{{\phi}}_{k_l}^\dag{\phi}_{k_h}=
\tilde{{\phi}}_{\bar{k}_l}^\dag{\phi}_{\bar{k}_h}=
\delta_{lh}\;{,}
\label{dual1}
\end{equation}
from which follows the completeness conditions
\begin{equation}
\sum^{M}_{l=1}
{\phi}_{k_l}\tilde{{\phi}}_{k_l}^\dag=
\sum^{M}_{l=1}{\phi}_{\bar{k}_l}\tilde{{\phi}}_{\bar{k}_l}^\dag=
{\cal I}\;{.}
\label{dual2}
\end{equation}

Equation (\ref{digf}) shows the
retarded Green function for a doubly infinite system.
For a semi-infinite lead, this must be modified to satisfy the
boundary conditions at the end of the leads.
Consider first the left lead, which extends to $z=-\infty$ and terminates
at $ z=z_0-1$, such that
the position of
the first missing slice is $z=z_0$.
To satisfy the boundary condition that the Green function must vanish at
$z=z_0$, we subtract from the right hand side of (\ref{digf}) a wavefunction
of the form

\begin{equation}
\Delta_z(z',z_0)=\sum^{M}_{lh}{\phi}_{\bar{k}_h}
e^{i\bar{k}_hz}
\Delta_{h{l}}(z',z_0)\;{,}
\label{delt1}
\end{equation}

where $\Delta_{h{l}}(z',z_0)$ is a complex matrix, determined from the
condition that the Green function vanishes at $z_0$, which
yields
\begin{eqnarray}
\nonumber
\Delta_z(z^\prime,z_0)=\Delta_{z^\prime}(z,z_0)=\;\;\;\;\;\;\;\;\;\;\;\;\;\; \\
\nonumber \\
\sum^{M}_{l,h=1}{\phi}_{\bar{k}_h}e^{i\bar{k}_{h}(z-z_0)}
\tilde{{\phi}}_{\bar{k}_h}^\dag{\phi}_{k_l}
e^{ik_l(z_0-z^\prime)}\tilde{{\phi}}_{{k}_l}^\dag{\cal{V}}^{-1}
\;{,}\nonumber \\
\label{delt}
\end{eqnarray}

For the purpose of computing the scattering matrix, we shall require the
Green function of the semi-infinite left-lead
$\tilde{g}_{zz^\prime}(z_0)=g_{zz^\prime}-\Delta_z(z^\prime,z_0)$ evaluated on
the surface of the lead, namely at $z=z'=z_0-1$.
Note that in contrast with the Green's function of a doubly infinite lead, which
depends only on the difference between $z$ and $z^\prime$, the Green's
function $\tilde{g}$ of a semi-infinite lead for arbitrary $z, z^\prime$
is also a function of the position $z_0$ of the first missing slice beyond the
termination point of the lead.
Writing $g_{\rm{L}}=g_{(z_0-1)(z_0-1)}(z_0)$ yields for this surface
Green function

\begin{equation}
g_{\rm{L}}=\left[{\cal I}-\sum_{l,h}
{\phi}_{\bar{k}_h}e^{-i\bar{k}_{h}}
\tilde{{\phi}}_{\bar{k}_h}^\dag{\phi}_{k_l}
e^{ik_l}\tilde{{\phi}}_{k_l}^\dag
\right]{\cal{V}}^{-1}\;{.}
\label{gleft}
\end{equation}

Similarly on the surface of the right lead, which extends to $z=+\infty$,
the corresponding surface Green function is

\begin{equation}
g_{\rm{R}}=\left[{\cal I}-\sum_{l,h}
{\phi}_{k_h}e^{ik_h}
\tilde{{\phi}}_{k_h}^\dag{\phi}_{\bar{k}_l}
e^{-i\bar{k}_l}
\tilde{{\phi}}_{\bar{k}_l}^\dag
\right]{\cal{V}}^{-1}\;{,}
\label{gright}
\end{equation}

The expressions (\ref{gleft}) and (\ref{gright}), when used in
conjunction with (\ref{hacca}) form a versatile method of determining
lead Green functions, without the need to perform k-space integrals
or a contour integration.
As a consequence of translational invariance of the doubly infinite system,
the surface
Green functions are independent of the position of the surface
$z_0$. 
Furthermore as noted
below, in the case of different vectors ${\phi}_{k}$ 
corresponding to the same real k-vector $k$, the current operator is not
diagonal. Hence it is convenient to perform a unitary rotation in such degenerate
sub-space to ensure the unitarity of the S-matrix.

\subsection{The effective Hamiltonian of the scattering region.}

Given the Hamiltonian of a scattering region and a matrix of couplings
to the surfaces of external leads, the Green function of the
scatterer plus leads can be computed via Dyson's equation.
For structures of the form of Fig 1, which possess a quasi-one dimensional
geometry and a Hamiltonian which is block tri-diagonal, this task can be
made more efficient by first projecting out the internal
degrees of freedom of the scatterer, to yield an effective Hamiltonian
involving only those degrees of freedom on the surfaces of
the external leads. In the literature, depending
on the context or details of implementation, this
procedure is sometimes referred to as \lq\lq the recursive Green function
technique" or \lq\lq the decimation method", but is no more than
an efficient implementation of Gaussian elimination.

Consider a scatterer composed on $N-2M$ degrees of freedom. Then the
Hamiltonian for the scatter plus semi-infinite leads is of the form
$H=H_L+H_R+\tilde H$, where $H_L$, $H_R$ are the Hamiltonians of the
left and right isolated leads and $\tilde H$ a $N{\rm x} N$ Hamiltonian
describing the scattering region and any additional couplings involving
surface sites of the leads induced by the presence of the scatterer.
The aim of the decimation (i.e. recursive Green function)
method is to successively eliminate the internal degrees of freedom of the scatterer,
which we label $i$, $i= 1,2,...,N-2M$, to yield a $(2M) {\rm x} (2M)$
effective Hamiltonian $H_{\rm eff}$.
After eliminating the degree of freedom $i=1$, $\tilde H$ is reduced to
a $(N-1) {\rm x} (N-1)$ matrix with elements

\begin{equation}
H_{ij}^{(1)}=\tilde H_{ij}+\frac{\tilde H_{i1}\tilde H_{1j}}{E-\tilde H_{11}}
\label{eig2}
\end{equation}

Repeating this procedure $l$ times we obtain the ``decimated''
Hamiltonian at $l$-th order

\begin{equation}
H_{ij}^{(l)}=H_{ij}^{(l-1)}+\frac{H_{il}^{(l-1)}H_{lj}^{(l-1)}}{E-H_{ll}^{(l-1)}}
\;{,}
\label{eigl}
\end{equation}

and after $N-2M$ such steps, an effective Hamiltonian $H_{\rm eff}=H^{N-2M}$
of the form

\begin{equation}
H_{\rm{eff}}(E)=\left(
\begin{array}[4]{rr}
H_{\rm{L}}^*(E) & H_{\rm{LR}}^*(E)\\
H_{\rm{RL}}^*(E) & H_{\rm{R}}^*(E)\\
\end{array}
\right){\;},
\label{haccasup}
\end{equation}

In this expression,
 $H_{\rm{L}}^*(E)$ ($H_{\rm{R}}^*(E)$) describes intra-surface
couplings involving degrees of freedom belonging to the surface of
the left (right) lead
 and $H_{\rm{LR}}^*(E)
=H_{\rm{LR}}^*(E)^{\dagger}$
describes the effective coupling
between the surfaces of the left and the right leads.

Since the effective Hamiltonian is energy dependent, this procedure
is particularly useful method when we wish to compute the Green function
at a given energy. It is also very efficient in the presence of short range
interactions, because
only  matrix elements involving degrees of freedom
coupled to the decimated one, are redefined.
Since the problem now involves only  $(2M) {\rm x} (2M)$ matrices,
it is straightforward to obtain the surface
Green function for the whole system
(i.e. the scattering region attached to semi-infinite leads)
by solving  Dyson's equation

\begin{equation}
G(E)=(g(E)^{-1}-{{H}}_{\rm{eff}}(E))^{-1}
\;{,}
\label{dys}
\end{equation}

where
\begin{equation}
g(E)=\left(
\begin{array}[4]{rr}
g_{\rm L}(E) & 0\;\;\;\;\;\; \\
0\;\;\;\;\;\; & g_{\rm R}(E) \\
\end{array}
\right){\;},
\label{ggg}
\end{equation}
with $g_L$ and $g_R$ given by equations (\ref{gleft}) and (\ref{gright}).

\subsection{The scattering matrix and transport coefficients}

To  extract transport coefficients
from the Green function, we generalize the method described in
\cite{lhr} (in particular see A.26 of \cite{lhr})
to the case of non-orthogonal scattering channels.
For a system of Hamiltonian $H$, the $S$ matrix
is defined to  connect incoming to outgoing
propagating states in the external leads.
If $k$, ($k'$) are real incoming (outgoing) wavevectors of energy $E$,
 then an incident plane-wave in one of the leads, with
longitudinal wavevector $k$, will scatter into outgoing plane-waves
$k'$ with amplitudes $s_{k'k}(E,H)$. If all plane-waves are
normalized to unit flux, (by dividing by the square-root of their
group velocities) then provided the plane-wave basis diagonalizes the
current operator in the leads, the outgoing flux along channel
$k'$ is $\vert s_{k'k}(E,H)\vert^2$ and $S$ will be unitary.
If $H$ is real, then $S$ will be symmetric, but more generally
time reversal symmetry implies $s_{k'k}(E,H)=s_{kk'}(E,H^*)$.
For convenience, if $k,k'$ belong to the left (right) lead, then we define
reflection coefficients via $r_{k'k}=s_{k'k}$ ($r'_{k'k}=s_{k'k}$),
whereas if $k,k'$ belong to left and right leads respectively
(right and left leads respectively) we define transmission coefficients
$t_{k'k}=s_{k'k}$ ($t'_{k'k}=s_{k'k}$).

To extract transport properties for
the system of Fig.\ref{greensch}, consider the probability
current for an electron in the Bloch state (\ref{jwf})

\begin{equation}
J_k=n_{k_\perp}v_{k_\perp}
\;{,}
\label{jj1}
\end{equation}
where $n_{k_\perp}$ is the probability of finding an electron 
in a slice and $v_{k_\perp}$ is the corresponding group velocity.
It follows that the vector 
\begin{equation}
\psi_{z}=\frac{1}{\sqrt{v_k}}e^{ikz}{\phi}_k
\;{,}
\label{jj4}
\end{equation}
is normalized to unit flux.
To compute the group velocity we note that if $|\psi_k>$
is an eigenstate (\ref{ham}), whose projection onto slice $z$ is
$\psi_z$, then
\begin{eqnarray}
\nonumber
v_k=\frac{1}{\hbar}\frac{\partial}{\partial k}<\psi_k|H|\psi_k>=
\;\;\;\;\;\;\;\;\;\;\; \\
\nonumber \\
=\frac{1}{\hbar}\frac{\partial}{\partial k}
\left[{\phi}_k^\dag\left(H_0+H_1e^{ik}+H_{-1}e^{-ik}
\right){\phi}_k\right]=
\;\; \\
\nonumber \\
=\frac{i}{\hbar}
{\phi}_k^\dag\left(H_1e^{ik}-H_{-1}e^{-ik}\right)
{\phi}_k
\;\;\;\;\;\;\;\;{,}\nonumber \\
\label{jj6}
\end{eqnarray} 
where the last step follows from equation (\ref{scsum2})
and normalization of $\phi_k$.

It can be shown that the states (\ref{jj4})
diagonalize the current operator
only if they correspond to distinct 
$k$ values. In the case of degenerate $k$'s, the current is in
general non-diagonal.
Nevertheless  it is always
possible to define a rotation in the degenerate subspace
for which the current operator is diagonal
and in what follows, when a degeneracy
is encountered, we assume that such a rotation has been performed.
With this convention, the current carried by a state of the form
\begin{equation}
\psi_z=\sum_l a_l\frac{e^{ik_lz}}{\sqrt{v_l}}{\phi}_{k_l}
\;{,}
\label{co5b}
\end{equation}
is simply $\sum_l \vert a_l\vert^2$.

It is now straightforward to generalize the analysis of \cite{lhr} to
the case of non-orthogonal scattering channels.
Consider first a doubly infinite periodic structure, whose
 Green function is given by
 equation (\ref{digf}). For $z\geq z'$,
acting on $g_{zz^\prime}$ from the right with
 the following projector

\begin{equation}
P_l(z^\prime)={\cal{V}}{\phi}_{k_l}\frac{e^{ik_lz^\prime}}
{\sqrt{v_l}}
\;{,}
\label{ss3}
\end{equation}
yields the normalized plane-wave (\ref{jj4}). Similarly by acting
on the Green function $g_{zz^\prime}(z_0)$ of a semi-infinite left-lead 
terminating
at $z_0$, one obtains for $z\geq z'$, $z_0\geq z$, an eigenstate of a
semi-infinite lead arising from a normalized incident wave along channel
$k_l$.

Thus the operator $P_l(z')$ and its left-going counterpart
$\bar{P}_l(z')$ allow us to project-out wavefunctions from the Green
function of a given structure.
For example if $G_{zz^\prime}$ is the retarded
Green function for a scattering region
sandwiched between two perfect leads whose surfaces are located
at the points $z=0$ and $z=L$,
then for
$z'\leq 0$, the projected wavefunction is of the form

\begin{equation}
\psi_z=\left\{
\begin{array}[4]{r}
\frac{e^{ik_lz}}{\sqrt{v_l}}{\phi}_{k_l}+
\sum_{h}\frac{r_{hl}}{\sqrt{\bar{v}_{h}}}
e^{i\bar{k}_{h}z}{\phi}_{\bar{k}_h}
\;\;\;\;\;z\leq{0} \\
\\
\;\;\;\;\ \sum_{h}\frac{t_{hl}}{\sqrt{{v}_{h}}}
e^{i{k}_{h}z}
{\phi}_{{k}_h}\;\;\;\;\;z\geq{L} \\
\end{array}
\right.\;
\;{,}
\label{uave2}
\end{equation}

where $r_{h{l}}=r_{\bar k_{h}, k_{l}}$ ,
$t_{h{l}}=t_{k_{h},k_{l}}$ are
 reflection  and transmission coefficients associated with an incoming state
 from the left.  In particular for $z=L$, $z'=0$, one obtains

\begin{equation}
\sum_{h}\frac{t_{h{l}}}{\sqrt{{v}_{h}}}
e^{i{k}_{h}L}{\phi}_{k_h}=G_{L0}P_l(0)
\;{,}
\label{ss9}
\end{equation}
and hence

\begin{equation}
t_{h{l}}=\tilde{{\phi}}_{{k}_h}^\dag
G_{L0}{\cal{V}}{\phi}_{k_l}
\sqrt{\frac{{v}_{h}}{{v}_{l}}}e^{-ik_{h}L}
\;{.}
\label{ss11}
\end{equation}

Since the right hand side of (\ref{ss11}) involves only
the surface Green function of equation (\ref{dys})
the transmission coefficients (and by analogy all other transport
coefficients) are determined.
Since the above analysis is valid for any choice of the Hamiltonians $H_0$
and $H_1$, this approach is very general.

\section{Results for Co and Ni based multilayers}

Using the technique developed above, we have 
studied transport properties of multilayers formed from
Co and Ni as magnetic materials and several 3$d$, 4$d$ and 5$d$ transition metals
as non-magnetic materials. All of these metals
possess an fcc lattice structure with the following lattice constants
(Tab.\ref{tavola1}).

It is clear that  Co and Ni have a good lattice match with Cu,
while for the other metals the lattice mismatch is large
and may introduce distortion and defects at
the interface. The latter introduces an additional scattering
at the interfaces, but is neglected in our calculations.
Nevertheless
we will show that large values of the GMR ratio can be obtained,
in agreement
with the largest  experimental values, which suggests that CPP GMR
is a bulk effect, whose main features are
contained in a ballistic
quantum description of 
the conductance with an accurate band structure.
In what follows we model all the metals by
an $spd$ tight-binding Hamiltonian with nearest neighbor couplings,
whose parameters are chosen to fit the band structure evaluated
from first principle calculations \cite{papaco}. The hopping parameters at the 
heterojunctions between different materials are assigned the geometric
mean of the pure metal values.

\subsection{The  DOS and Conductance of the pure metals}

We begin our analysis by examining the DOS and  conductance of the
pure metals. Since the Hamiltonians
include $spd$ hybridization, angular momentum states are not
eigenstates of the system. 
Nevertheless  to understand
the relative r\^ole of the angular momentum 
inter-band and intra-band scattering, it is useful to project
the DOS and conductance onto an angular momentum basis.
We will label as an $s$-like electron (and similarly for the $p$ and $d$ electrons)
an electron whose $s$-component $|<s|\psi>|^2$ of the wave function
$|\psi>$ is much larger than the $p$ and $d$ components. 

The DOS's for the two spin sub-bands of Co and Ni are
very similar and as an example the Co DOS is shown in Fig.\ref{codos}.

As in all
the $d$ transition metals, the DOS is formed from a
localized $d$ band embedded in parabolic $s$ and $p$ bands. The broadening of the
bands is roughly the same in Co and Ni, as well as the position
of the majority band
with respect to the Fermi energy. In both materials,
the Fermi energy lies
just above the edge of the majority $d$ band,
while the minority band is obtained from a rigid shift
of the majority band toward higher energies,  the magnitude of which
is larger in Co than in Ni.
In both the minority bands of Co and Ni the Fermi energy lies well 
within the $d$ band and the DOS is completely dominated by the $d$ electrons. 
A rough estimate of the mismatch between the minority $d$ bands of Co and Ni can
be obtained from the on-site energies of the $d$ electrons in the
minority band. As shown in table \ref{tavazza},
the difference between the on-site energies of the $d$ minority
electrons in Co and Ni is about 0.7eV and corresponds to the relative shift
of the bands (the on-site energies shown in the table are
chosen in order to have the Fermi energy $E_{\rm F}=0$).

The  conductance of pure Co and Ni is determined solely by the DOS.
For majority electrons at the Fermi
energy, the current is carried by the $s$, $p$ and $d$ electrons,
which give almost equal contributions.
 On the other hand the current carried by minority electrons is completely
dominated by the $d$ electrons, with the contributions from $s$ and $p$ electrons
being no larger than 10\%.
If we neglect the relative shift in energies of the
minority bands, the Ni and Co conductances possess the same qualitative features
and since the effective mass is proportional
to the inverse of the band width, we find that
the current carried by majority electrons is formed from a mixture
of light $s$ and $p$ electrons and heavy $d$ electrons, whereas
the minority-electron
 current is carried almost entirely by heavy $d$ electrons.

Now consider the non-magnetic
3$d$, 4$d$ and 5$d$ transition metals with fcc lattices. A glance at
the DOS of these materials reveals three types of band structure:
i) the DOS closely matches the DOS of the majority
spin sub-band of Co and Ni (e.g. Cu and Au),
ii) the DOS has only $sp$ components at the Fermi energy, with the d
component highly suppressed (as in Ag), iii) the DOS is composed of an
 almost pure
 $d$ component at the Fermi energy (e.g. for Pd and Pt).
 Examples of each of these are given in  figure \ref{cuagpd}, which shows
the DOS of Cu, Ag and Pd and in figure \ref{curcuagpd}, which shows the
corresponding conductances.

In what follows, we shall show that for ballistic structures,
the mismatch between the bands of the magnetic
and non-magnetic metals forming the multilayer is the key feature
which determines the conductance. Moreover,
although the positions of the $s$ and $p$ bands are the same for both spins,
for $d$-electrons the two spin sub-bands possess a different
mismatch at the interface and therefore this mismatch 
(potential step for $d$ electrons at the interface)
largely determines the magnitude of CPP GMR.

In addition to the above generic features,
the on-site energies
of the $s$ bands of Au and Pt are small compared with those
 of Ni and Co. In the Co/Au and Co/Pt multilayers,
this may induce
 strong scattering of the $s$-electrons, resulting in a
 strong suppression of the $s$-electron contribution to the
total conductance.
This effect should be weaker in the
Pt-based than in the Au-based multilayers, because
 DOS of Platinum has mainly a $d$ character at the Fermi energy,
 which means that
 the Co/Pt 
multilayers will feel the effect of the large $s$-band mismatch only through the
hybridization of the $s$ and $d$ bands at the Fermi energy.

Figure \ref{curcuagpd}
shows how these three distinct DOS characteristics are reflected
in the conductances of the normal metals and give rise to
three different scenarios for charge transport:
i) the contributions to the current
from $s$, $p$ and $d$ electrons are almost equal
(e.g. in Cu and Au), ii) the current has a strong $sp$ character (e.g. in Ag),
iii) the current
has a strong $d$ character (e.g. Pd and Pt).

These different characteristics of the current carriers
in the non-magnetic metals give rise to another important source of
interface scattering.
Since the majority spins in the magnetic metals are mainly
$sp$ electrons with light effective masses and the minority spins are
d electrons with heavy effective masses, it is clear that depending on the
choice of
non-magnetic metal, different spin-dependent inter-band scattering must occur at
the interfaces.
For example in the Co/Ag system,
a majority spin propagates in Co as a mixture of $s$, $p$ and $d$ electrons
and in Ag as an $sp$ electron.
This means that an
electron in the Ag, whose spin is in the same direction of the magnetization,
can enter Co as an $sp$ electron without the need for strong inter-band
scattering. On the other hand if its spin points in the opposite direction, 
it will undergo inter-band scattering, because
in the minority band the electron must propagate as a $d$ electron.
Moreover the inter-band scattering involves final states
with a large DOS, and hence a scattering is expected to be strong.

The above observations suggest that the
key mechanisms affecting transport are
i) a strong band mismatch and ii) a strong
inter-band scattering. The best GMR multilayers must be able to maximize
electron propagation in one of the two spin bands and to minimize it in
the other. To achieve this result the high conduction spin band should have a
small band mismatch and weak inter-band scattering at the heterojunctions, while
the
low conduction band should have a large band mismatch and strong inter-band
scattering.

\subsection{A comparison between Co-based and Ni-based multilayers.}

To clarify how the spin polarization
of the magnetic material affects the properties of the GMR multilayers,
we begin by examining  GMR in Cu-based multilayers,
 in which the magnetic metals are either Ni or Co.
All the multilayers consist of ten bilayers of the form
A/Cu where A is Co or Ni, attached to two semi-infinite Cu leads.
After calculating the different spin conductances in the ferromagnetic
and antiferromagnetic configurations, the GMR ratio is obtained from equation
(\ref{gmr}).
In all our calculations  the current flows in the
(110) crystalline direction and the structures are translationally invariant
within the layers.
In what follows we consider $8100$ $k_\parallel$ points
($90\times90$) in the plane of the layers. We have estimated that the GMR ratio
calculated with $2\times 10^4$ $k_\parallel$ points
on average differs by $\sim3\%$, from
that calculated using $8100$ $k_\parallel$ points
($\frac{{\rm GMR}(8100)-{\rm GMR}(2\times 10^4)}
{{\rm GMR}(8100)}\sim3\%$). Since the
oscillations of the GMR ratio with respect to the layer thicknesses
are larger than $3\%$, the choice of $8100$ $k_\parallel$ points
allows us
to investigate the oscillating behavior of the conductance and the GMR,
and is a good compromise between the accuracy of the calculation and the 
required computer time. 

Initially we fix
the magnetic layer thickness to 5 atomic planes (AP), 
and calculate the conductance and GMR as a function of the Cu layer thickness.
In what follows, we normalize the conductance by dividing
by the conductance of a single spin
in the pure metallic leads (hence in this
case by one half of the total Cu conductance, because of
spin degeneracy), to yield the results shown in  figure \ref{conigmr}.

{}From  figure \ref{conigmr} it is clear that the Co based multilayers possess
larger GMR ratios. In the ferromagnetic configuration,
the majority electrons possess high conductances in both cases, reflecting
the good match between the majority bands of Co and Ni, and the Cu band.
Moreover the better match of the $s$ and $p$ majority bands of Ni with Cu,
compared with those of Co, gives
rise to a slightly higher conductance in majority channel for Ni than for Co.
A similar argument explains the difference in the conductances of the minority
channel.
As we can see from table \ref{tavazza},
the minority $d$ band of Ni is a better match to Cu than that  of Co, as indicated
by the difference in the on-site energies about 0.7 eV.
Hence for the minority band, the interface scattering between Co/Cu
is greater for Ni/Cu.
In the antiferromagnetic configuration,
both spins undergo  the same scattering sequence, belonging alternately
to the majority and to the minority bands. The total spin conductance in the 
antiferromagnetic configuration is found to be close
to that of the minority
band in the ferromagnetic configuration, because
the minority band mismatch is larger than the majority band,
and dominates the scattering.

The ratio $R$ between the conductance ($\Gamma$) of the AF configuration and 
of the minority band in the F configuration ($R=\Gamma({\rm AF})/
\Gamma({\rm F}\:{\rm minority}))$
is $\sim0.6$ for Co/Cu and $\sim0.9$ for Ni/Cu. 
This difference can be understood by modeling the interface scattering
through an effective step potential, whose magnitude is equal
to the band mismatch, as will be discussed in a future publication.
The effective scattering
potential in the antiferromagnetic configuration will be a sequence of 
high steps (minority band) and low
steps (majority band). The calculated $R$ ratios arise, because
the perturbation of the minority steps due to the majority steps,
is smaller in Ni/Cu than in Co/Cu.
{}From this analysis the splitting between the two spin sub-bands in the
magnetic materials
is the crucial parameter leading to large GMR ratios and since
such splitting is larger in Co than in Ni, Co emerges as a natural candidate
for high GMR ratio multilayers. Note that highest possible values of
GMR can probably be achieved with the use of half-metallic
ferromagnets with 100\% spin polarization of electrons \cite{alex1}.

Having examined the dependence of transport properties on the
normal-metal layer thickness, we now examine the dependence on the
magnetic-layer thickness. For a fixed Cu layer-thickness of 5 atomic
planes, figure \ref{magvar} shows results for Co/Cu and Ni/Cu
multilayers.

A key result of this figure is that
for thin magnetic layers,
GMR in both Ni/Cu and Co/Cu multilayers
is suppressed due to tunneling through the effective potential barrier.
Thus we predict a lower limit of approximately 4 atomic planes to
the magnetic-layer thickness, in order to achieve the
highest possible GMR ratio.
In what follows we will only consider thicknesses
larger than this value.

\subsection{Dependence of GMR on
non-magnetic spacer material.}

We now consider the dependence of GMR on the choice
of non-magnetic material
in Co and Ni-based multilayers. In all calculations we fix
the Co thickness at 5 and 10 atomic planes and  vary the
thickness of the non-magnetic layers from 1 to 40 atomic planes. The
material in the external leads is the same non-magnetic material used for the
multilayers (e.g. Ag in Co/Ag multilayers).
Table \ref{tavola3} shows
the average value of the GMR ratio and the root mean square amplitude of oscillation
around such value ($\Delta$). To highlight the fact that GMR is an oscillatory
function of the normal-metal thickness with an amplitude which decreases
with increasing thickness, the table also shows
 the mean square oscillation calculated for
non-magnetic metal layers thicknesses between 1-10 ($\Delta1$).
In the table the subscript at Co indicates the number of atomic planes
of the Co layers.
The penultimate row of the table shows results for the Ni/Cu
system, for which we believe that no CPP experimental results are currently
available, even though the CIP conductance in high
magnetic field has been studied \cite{sato}.
We also speculate on the possibility
of using Co/Ni \cite{shul1,shul2} as a GMR material, 
results for which are shown on the
last row.

{}From the Table III it is clear that the GMR ratio results to depend quite
sensitively on the multilayer geometry, i.e. on the layer thicknesses.
In fact the simulations with the Co thickness fixed at 5 atomic planes seem
to suggest that the $sp$ conductors as spacer layers (Cu, Ag) result in
larger GMR ratios in this case. 
The simulations with 10 Co atomic planes show that
the multilayers with $d$-electron spacers (Pd, Pt) correspond to
relatively larger GMR.
Nevertheless it is clear that the Co based multilayers present much
larger GMR ratios than the Ni based multilayers.
The table also demonstrates that the conductors dominated by $d$ electrons,
namely Pd and Pt, possess  very
similar GMR ratios and amplitudes of oscillation and that
 Au possesses the  largest amplitude oscillations.
As examples, figure \ref{gmrgmr} shows plots of
the GMR ratio as a function of
the non-magnetic metal layer thickness for the Co/Ag and Co/Pd systems.

In all cases (excluding Au)
the oscillations are small compared with the average value of the
GMR ratio, suggesting that there is an additional contribution
to the long range oscillations observed
experimentally. This is most likely to arise from
 a periodic deviation
 from a perfect antiferromagnetic configuration, the possibility of which
 is neglected in our calculations.
It is important to point out that perfect antiferromagnetic
alignment of the multilayer in zero magnetic field is a
consequence of the exchange coupling of the adjacent magnetic
layers through the non-magnetic layer.
The strength and phase of such coupling depend critically
on the Fermi surface of the non-magnetic metal \cite{bruno}.
To the best of our knowledge no experimental data are available
for the $d$ conductor multilayers, for which the antiferromagnetic
configuration may not be achievable.
Nevertheless, in spin valve systems such an antiferromagnetic configuration
can be always obtained by tuning the coercive fields of the different
magnetic layers, for instance by an appropriate choice
of the spin valve geometry, or by using some magnetization pinning
technique. Hence our theoretical predictions on Co/Pd and Co/Pt
multilayers can, in principle, be tested in the spin valves.

The above results for the GMR ratio hide the material dependence of the
electrical conductance and with a view to comparing these with their band
structures, we now present results for
the conductances of the different 
spin channels and of the AF configuration.
In the tables \ref{tavola4}, \ref{tavola5} and \ref{tavola6}
we present the conductance
($\Gamma$), 
the mean conductance oscillation ($\Delta\Gamma$), their ratio
($\Delta\Gamma/\Gamma$),
the maximum of the conductance oscillations ($\Delta\Gamma_{\rm max}$) and
its ratio with the mean conductance (${\Delta\Gamma_{\rm max}}/{\Gamma}$),
respectively for the majority electrons in the ferromagnetic configuration, the minority 
electrons in the ferromagnetic configuration, and both  spins in the
antiferromagnetic configuration. All conductances are normalized to the
single-spin conductance of the non-magnetic-metal leads.

The Tables IV-VI illustrate that, with the exception of Au,
materials belonging to the same class 
have similar normalized conductances. For Cu and Ag the
majority (minority) band is a high (low) transmission band,
leading to a large GMR ratio for such materials.
The majority bands of these two materials
match that of Co and there is little interband scattering
(even less in Ag where the electrons at the Fermi
energy are completely $sp$).
In contrast the minority bands are subject to a  large scattering
potential due to the difference between the on-site
energies of the $d$-band, and also
large interband scattering, due to the full $d$ character of the minority $d$ band
of Co. On the other hand, for Pd and Pt, which are $d$ conductors,
both sub-bands undergo to high scattering
albeit for different reasons. The on-site energies
of the majority band of Co, and of Pd and Pt, are roughly the same, ensuring a 
good band match. Nevertheless the broadening of the $d$ majority band of Co
is associated with a mixing of $s$, $p$ and $d$ electrons, 
while the Pd and Pt bands are mainly $d$-like. Hence, in the majority band 
of Co/Pd and Co/Pt superlattices,
large inter-band scattering is present.
In contrast the minority band is $d$-like in Co, Pd, and Pt, but
there is a significant
 difference in the on-site energies, resulting in a
large effective potential at the interface.

The Au/Co multilayers lie somewhat outside the above picture,
because even though the $d$ band resembles that of Ag,
the on-site energies of the $s$ and $p$ bands are considerably
smaller than the corresponding bands in Co.
This means that  strong scattering occurs
in the $s$ and $p$ bands and since  the $s$ and $p$ electrons of Au
carry most of the current, there is a strong suppression
of the conductance in all spin channels. From the above tables,
we see
that the Co/Au system possesses a low conductance in all the spin
channels and in the antiferromagnetic configuration.

Finally we note that compared with  the majority spin channel
the  oscillations are larger in the minority spin channel and
in the antiferromagnetic configuration. This suppression of oscillations
in the former occurs because of
the better band matching in the majority band.
In figure \ref{conosc} we show the
conductance of the minority band as a function of the non-magnetic layers 
thickness, for all the materials studied.

The oscillations that we observe
 never exceed 20\% of their mean value (except
for the Co/Au system) and they are larger for smaller thicknesses.
This is substantially smaller then the observed values for Co/Ni 
system \cite{shul1,shul2}. The difference may in part be related
to scattering on disorder, always present in experimental systems,
and to simplicity of a tight-binding model that we used. In addition,
actual magnetic configuration in those systems can be somewhat
different as compared to an ideal one that we considered.
It is interesting to note that, generally, materials with
small conductances ($\Gamma<0.25$ using the usual normalization
for conductances) 
possess larger oscillations,
because low conductances indicate strong
scattering potentials, and hence larger fluctuations.
A qualitative picture of
 these conductance oscillations will be presented in a future publication,
 \cite{noi}
 where  the above quantitative results are compared
 with a simple Kronig-Penney model.

\section{Conclusion}

Firstly, we have developed a completely general Green's function
technique for elastic spin-dependent transport calculations, which
(i) scales linearly with a system size and 
(ii) allows straightforward application
to general tight-binding ($spd$ in the present work) Hamiltonians. 
This technique can be applied to
unrestricted studies of different systems, including tunneling spin
valves \cite{alex1,alex2} and magnetic multilayers with superconducting leads.
The formul\ae (\ref{gleft}) and (\ref{gright}) for the surface
Green functions of external leads are the central result of the first
part. Explicit general expressions for the Green's functions enable us to avoid
using a small imaginary part in energy.

Secondly,  we have presented an extensive study of transport
in magnetic multilayers in CPP (current perpendicular to planes)
geometry in the limit of large coherence length.
Ni and Co were considered as materials for magnetic
layers and several 3$d$, 4$d$ and 5$d$ metals as non-magnetic spacers.
Key parameters have been  identified as controlling a giant
magnetoresistance in those systems. These are the character of
electronic states at the Fermi level and a mismatch in
relevant band edges across interfaces.
We have found that, in accordance with experiment \cite{shul1,shul2}, there are
oscillations in the conductance as a function of both magnetic and spacer layer
thicknesses. The magnitude of the calculated oscillations is, however,
smaller than observed experimentally. 
Some reasons for this behavior have been indicated and deserve further study.
A semi-quantitative analysis of the oscillations will presented 
elsewhere\cite{noi}.

\section{Acknowledgements}

The authors wish to acknowledge Prof. Ivan Schuller for valuable
and useful discussions.
This work is supported by the EPSRC and the EU TMR Programme.

\newpage

\begin{figure}[h]
\begin{center}
\leavevmode
\epsfxsize=80mm \epsfbox{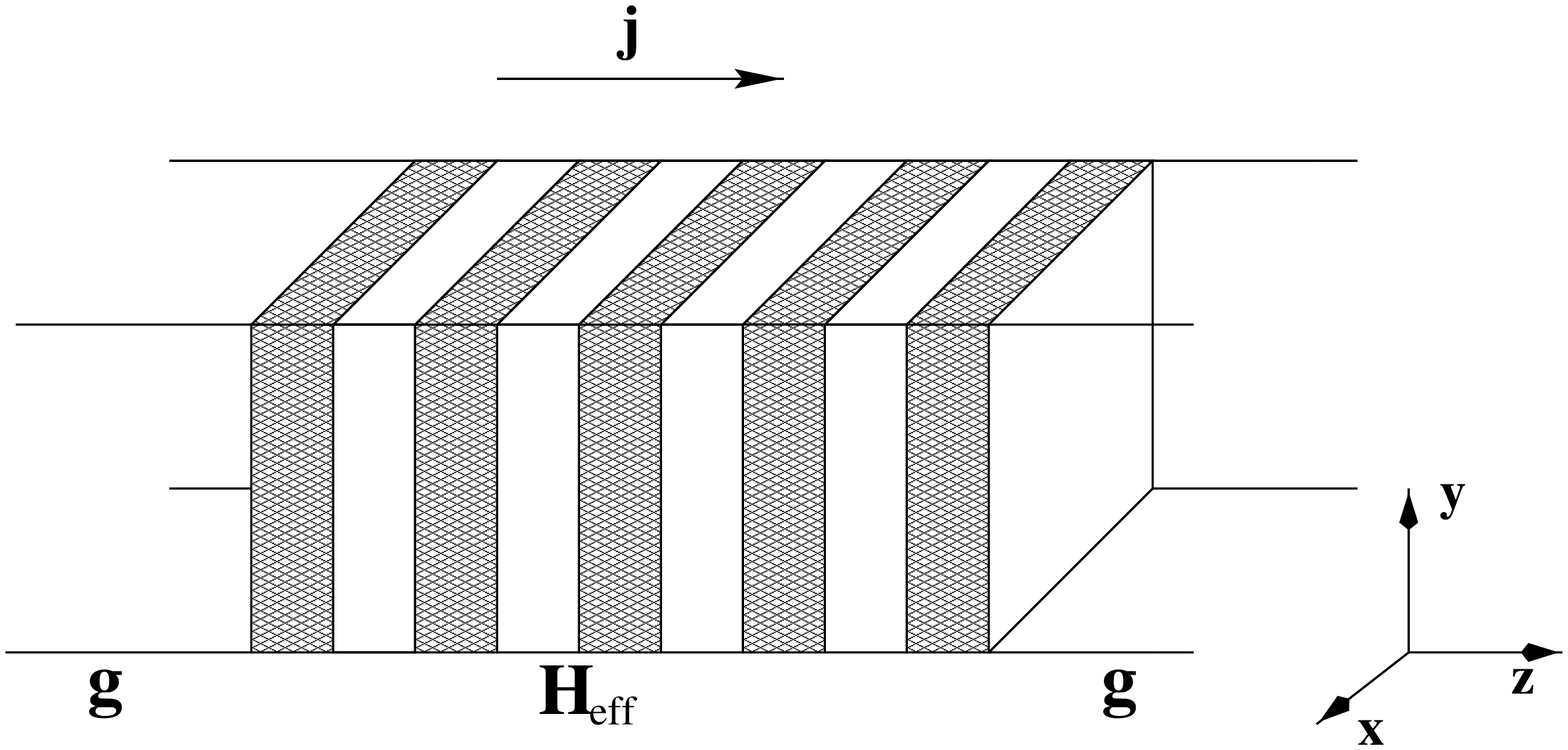}
\end{center}
\caption{Sketch of a finite superlattice connected to two semi-infinite leads.}
\label{twoprobe}
\end{figure}

\newpage

\begin{figure}[h]
\begin{center}
\leavevmode
\epsfxsize=80mm \epsfbox{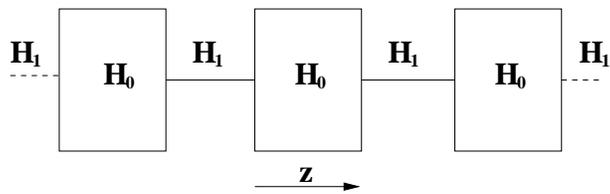}
\end{center}
\caption{A doubly infinite system formed from periodically repeated slices.}
\label{greensch}
\end{figure}

\newpage

\begin{figure}[h]
\begin{center}
\leavevmode
\epsfxsize=100mm \epsfbox{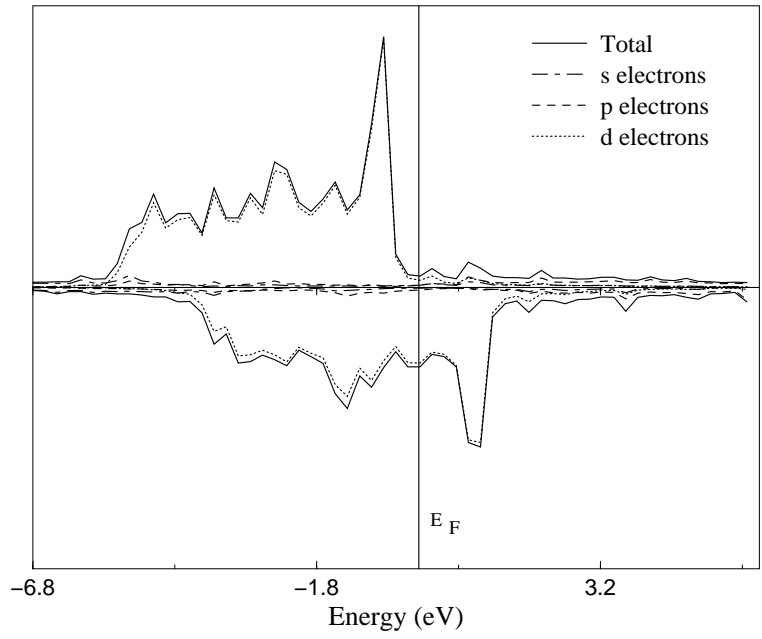}
\end{center}
\caption{DOS for pure Co. The vertical line denotes the position of $E_F$
that is chosen to be 0 eV.}
\label{codos}
\end{figure}

\newpage

\begin{figure}[h]
\begin{center}
\leavevmode
\epsfxsize=150mm \epsfbox{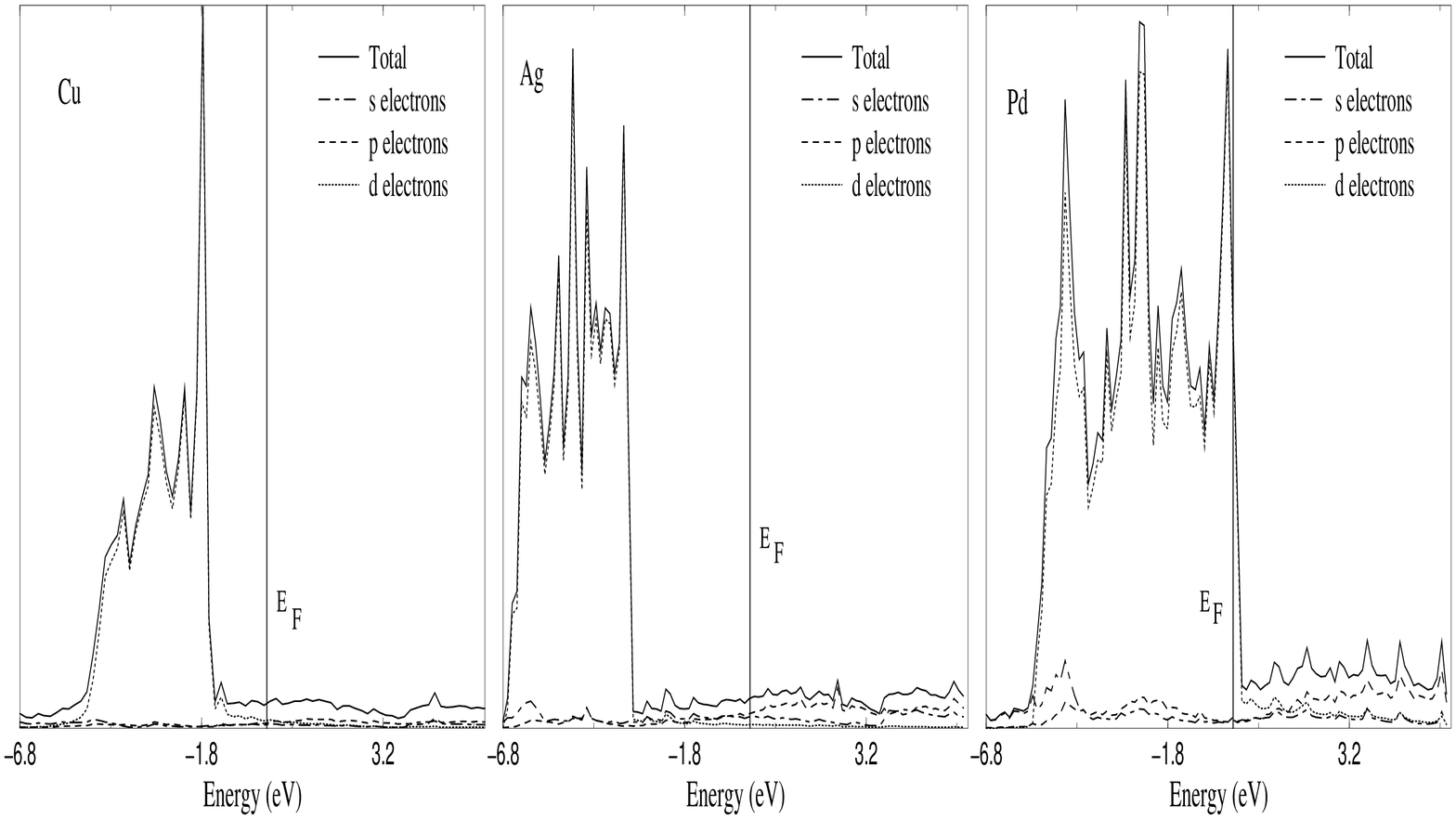}
\end{center}
\caption{DOS for pure Cu, Ag and Pd. The vertical lines denote the position
of the Fermi energy, which is chosen to be $E_F=0$}
\label{cuagpd}
\end{figure}

\newpage

\begin{figure}[h]
\begin{center}
\leavevmode
\epsfxsize=150mm \epsfbox{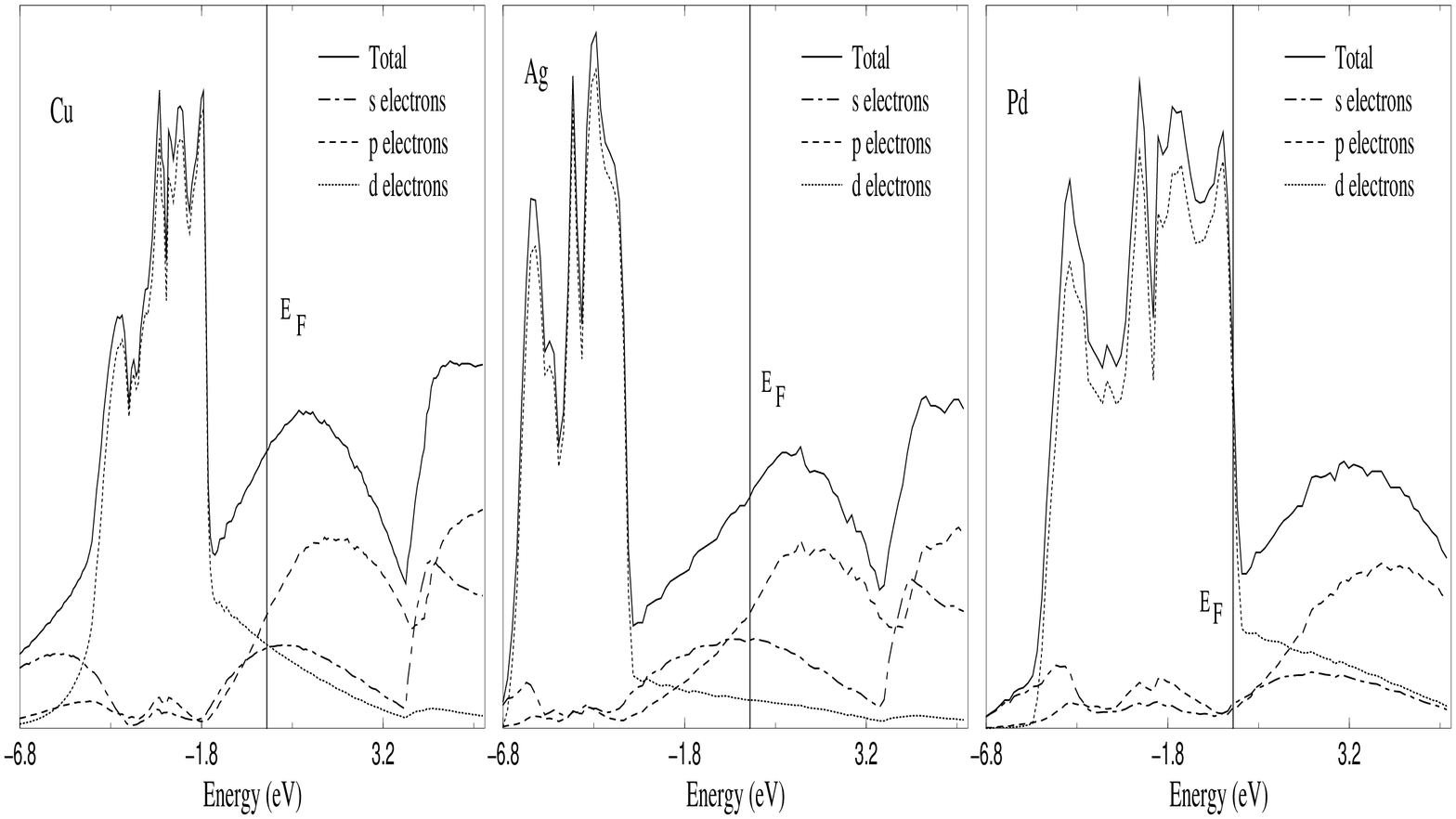}
\end{center}
\caption{The conductance for pure Cu, Ag and Pd. The vertical lines denote
the  the Fermi energy, $E_F= 0$.}
\label{curcuagpd}
\end{figure}

\newpage

\begin{figure}[h]
\begin{center}
\leavevmode
\epsfxsize=150mm \epsfbox{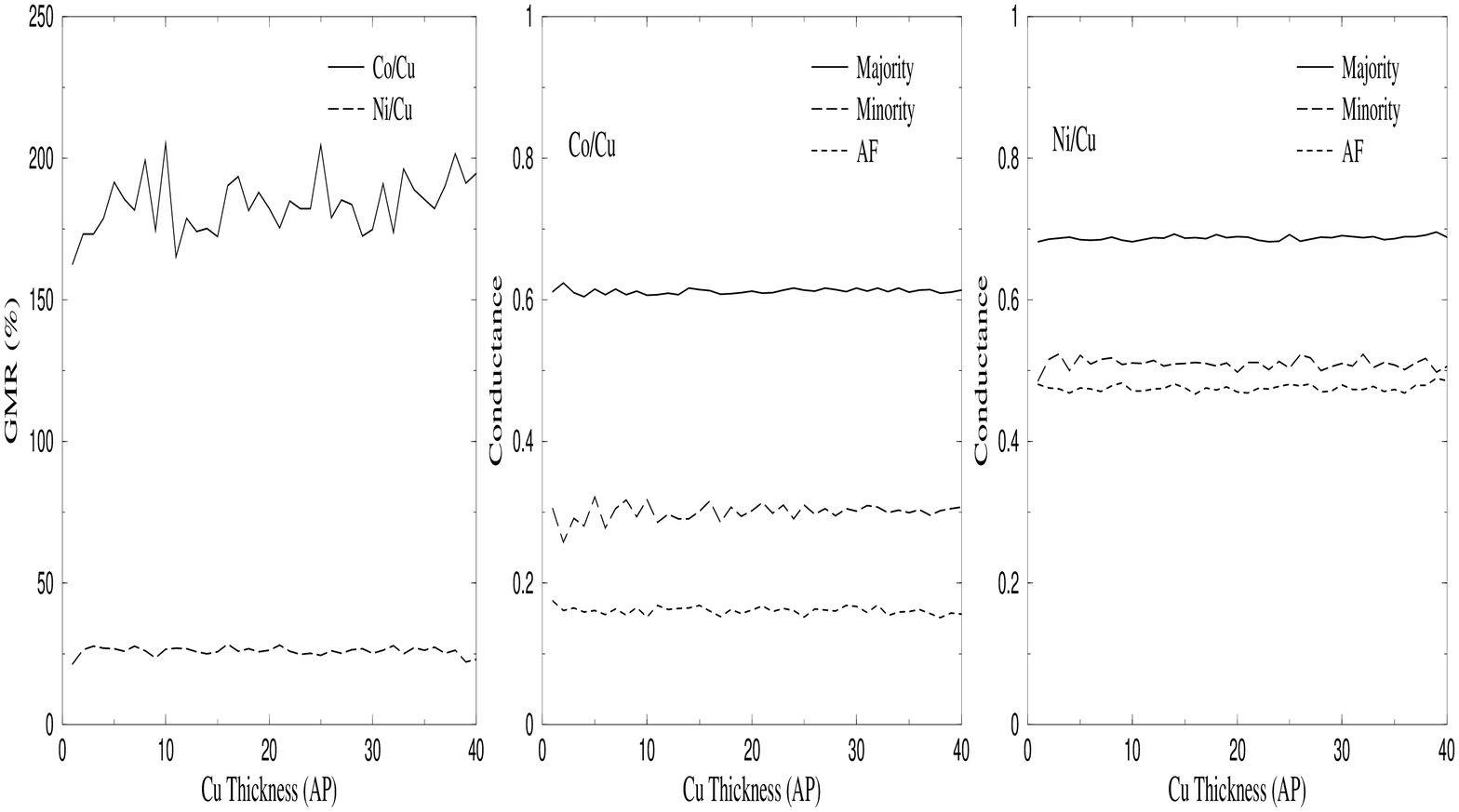}
\end{center}
\caption{GMR and spin conductance for Co/Cu and Ni/Cu systems 
as a function of the Cu layers thickness. The first graph
is the GMR, the second is the conductance for the Co/Cu system normalized to the 
conductance of pure Cu and the third is the conductance of the Ni/Cu system 
with the same normalization}
\label{conigmr}
\end{figure}

\newpage

\begin{figure}[h]
\begin{center}
\leavevmode
\epsfxsize=150mm \epsfbox{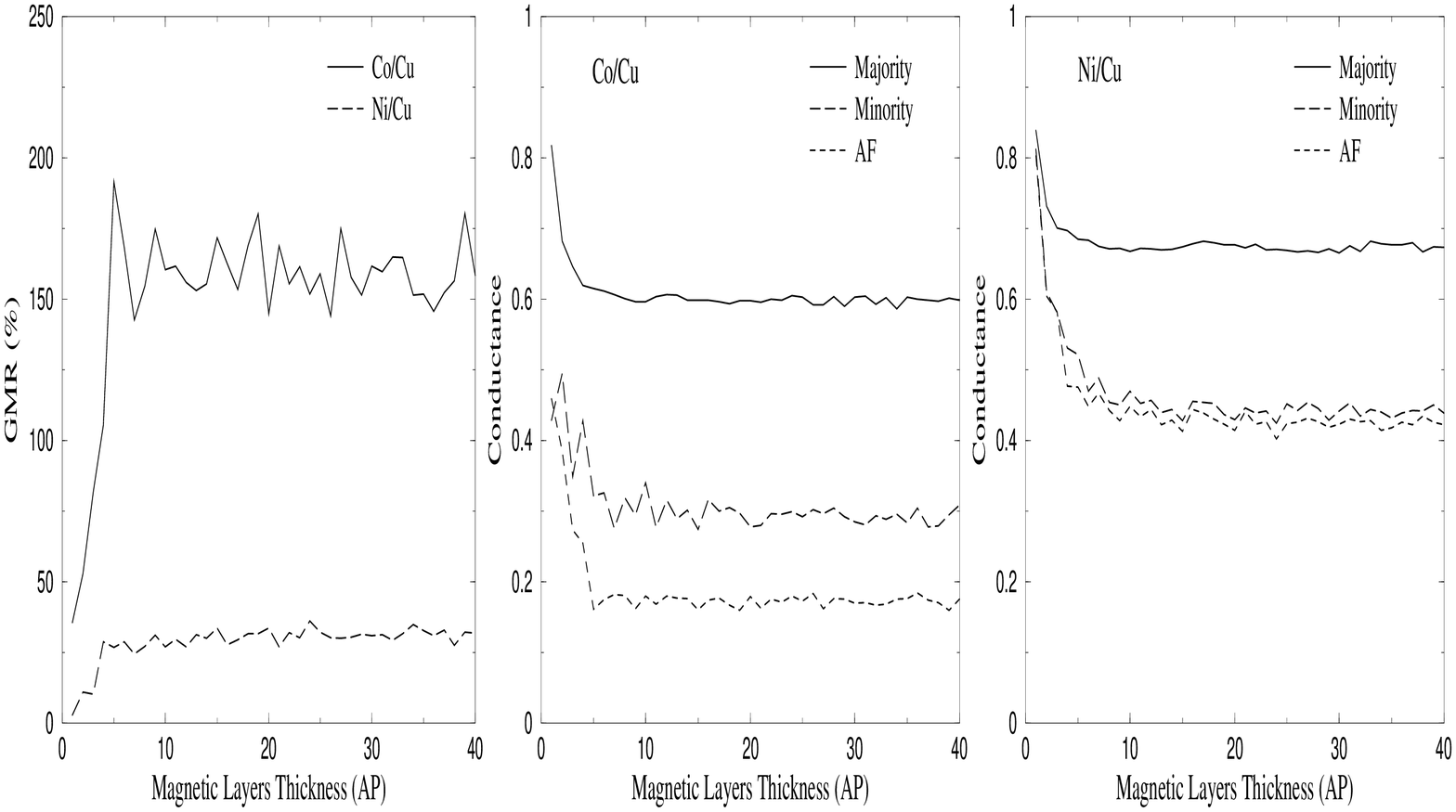}
\end{center}
\caption{GMR and spin conductance for Co/Cu and Ni/Cu systems 
as a function of Co and Ni layers thicknesses. The first graph
is the GMR, the second is the conductance for the Co/Cu system normalized to the 
conductance of pure Cu and the third is the conductance of the Ni/Cu system 
with the same normalization.}
\label{magvar}
\end{figure}

\newpage

\begin{figure}[h]
\begin{center}
\leavevmode
\epsfxsize=90mm \epsfbox{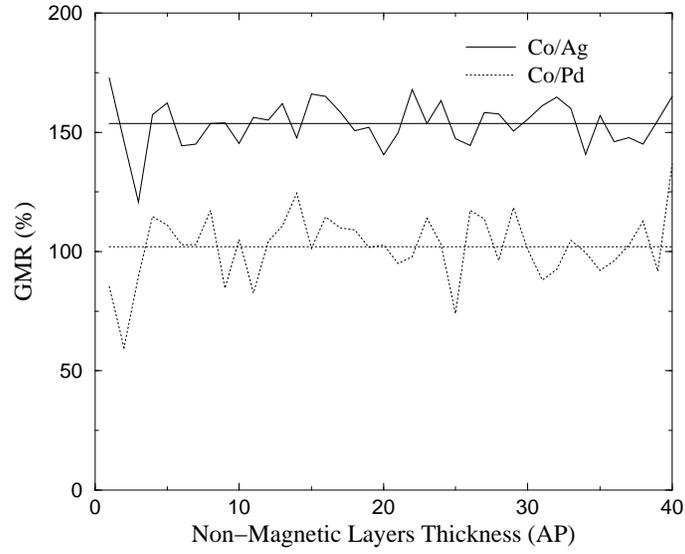}
\end{center}
\caption{GMR as a function of the non-magnetic metal layer
thickness for Co/Ag and Co/Pd. The horizontal lines denote the position of
the average GMR}
\label{gmrgmr}
\end{figure}

\newpage

\begin{figure}[h]
\begin{center}
\leavevmode
\epsfxsize=100mm \epsfbox{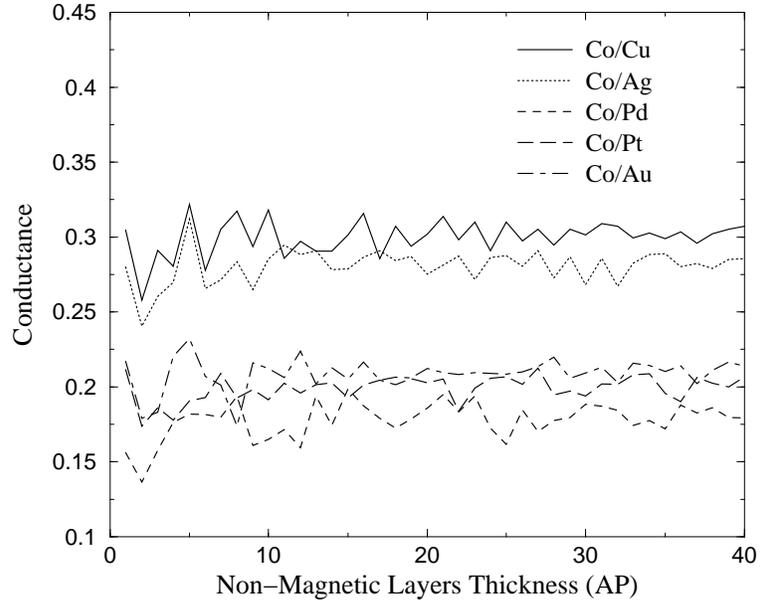}
\end{center}
\caption{Conductance of the minority spin electrons as a function of the 
non-magnetic layers thickness}
\label{conosc}
\end{figure}

\newpage

\begin{table}[htbp]
\begin{center}
\begin{tabular}{|c|c|} \hline
\ Metal \ & \ \ \ Lattice Constant (\AA) \ \ \ \\ \hline
{\bf Co} & 3.55 \\  \hline
{\bf Ni} & 3.52 \\  \hline
{\bf Cu} & 3.61 \\  \hline
{\bf Ag} & 4.09 \\  \hline
{\bf Pd} & 3.89 \\  \hline
{\bf Au} & 4.08 \\  \hline
{\bf Pt} & 3.92 \\  \hline
\end{tabular}
\caption{Lattice constants of the metals considered in the calculation}
\label{tavola1}
\end{center}
\end{table} 

\newpage

\begin{center}
\begin{table}
\begin{tabular}{|c|c|c|c|c|} \hline
\ Metal \ & \ \ \ $E_s$ (eV) \ \ \ & \ \ \ $E_p$ (eV) \ \ \ & \ \ \ $E_{d \: {\rm majority}}$ (eV) \ \ \ & \ \ \ $E_{d \: {\rm minority}}$ (eV) \ \ \ \\ \hline
{\bf Co} & 5.551 & 14.025 & -2.230 & -0.660\\  \hline
{\bf Ni} & 4.735 & 11.850 & -2.114 & -1.374\\  \hline
{\bf Cu} & 2.992 & 10.594 & -2.746 & -2.746\\  \hline
{\bf Ag} & 2.986 & 9.127 & -4.650 & -4.650\\  \hline
{\bf Pd} & 5.764 & 11.457 & -2.050 & -2.050\\  \hline
{\bf Au} & 0.329 & 10.081 & -3.823 & -3.823\\  \hline
{\bf Pt} & 1.849 & 11.523 & -2.614 & -2.614\\  \hline
\end{tabular}
\caption{On-site energies used in the calculations}
\label{tavazza}
\end{table}
\end{center}

\newpage

\begin{center}
\begin{table}[h]
\begin{tabular}{|c|c|c|c|c|c|} \hline
\ Multilayer \ & \ \ \ GMR ratio (\%) \ \ \ & \ \ \ $\Delta$ (\%) \ \ \ &
 \ \ \  $\Delta1$ (\%) \ \ \ & \ \ \ $\Delta$/GMR (\%) \ \ \ & \ \ \ $\Delta1$/GMR (\%) \ \ \ \\ \hline
{\bf Co$_5$/Cu} & 183.7 & 10.0 & 12.4 & 5.4 & 6.7 \\ \hline
{\bf Co$_5$/Ag} & 153.7 & 9.5 & 13.1 & 6.1 & 8.5 \\ \hline
{\bf Co$_5$/Pd} & 102.0 & 13.9 & 16.7 & 13.7 & 13.4 \\ \hline
{\bf Co$_5$/Pt} & 104.1 & 10.9 & 15.6 & 10.5 & 15.0 \\ \hline
{\bf Co$_5$/Au} & 98.8 & 20.4 & 33.62 & 20.6 & 34.0 \\ \hline
{\bf Co$_{10}$/Cu} & 150.7 & 9.2 & 9.2 & 6.1 & 6.1 \\ \hline
{\bf Co$_{10}$/Ag} & 131.0 & 7.6 & 5.3 & 5.8 & 4.1 \\ \hline
{\bf Co$_{10}$/Pd} & 165.2 & 31.1 & 32.2 & 18.8 & 19.4 \\ \hline
{\bf Co$_{10}$/Pt} & 175.7 & 14.8 & 21.1 & 8.4 & 12.5 \\ \hline
{\bf Co$_{10}$/Au} & 138.8 & 20.1 & 26.4 & 14.5 & 17.8 \\ \hline
{\bf Ni$_5$/Cu} & 25.9 & 1.5 & 1.8 & 5.8 & 6.9 \\ \hline
{\bf Ni$_5$/Co$_5$} & 66.1 & 4.1 & 6.6 & 6.2 & 10.0 \\ \hline
\end{tabular}
\caption{GMR ratio and GMR oscillations for different metallic multilayers}
\label{tavola3}
\end{table}
\end{center} 

\newpage

\begin{center}
\begin{table}[h]
\begin{tabular}{|c|c|c|c|c|c|} \hline
\ Multilayer \ & \ \ \ $\Gamma$  \ \ \ & \ \ \ $\Delta\Gamma$  \ \ \ &
 \ \ \  ${\Delta\Gamma}/{\Gamma}$ (\%) \ \ \ & \ \ \ $\Delta\Gamma_{\rm max}$
 \ \ \ & \ \ \  ${\Delta\Gamma_{\rm max}}/{\Gamma}$ (\%) \ \ \ \\ \hline
{\bf Co$_{5}$/Cu} & 0.61 & $3.76\cdot10^{-3}$ & 0.61 & $1.17\cdot10^{-2}$ & 1.92 \\ \hline
{\bf Co$_{5}$/Ag} & 0.66 & $4.10\cdot10^{-3}$ & 0.62 & $1.24\cdot10^{-2}$ & 1.88 \\ \hline
{\bf Co$_{5}$/Pd} & 0.35 & $5.32\cdot10^{-3}$ & 1.50 & $1.52\cdot10^{-2}$ & 4.29 \\ \hline
{\bf Co$_{5}$/Pt} & 0.38 & $5.01\cdot10^{-3}$ & 1.31 & $1.87\cdot10^{-2}$ & 4.91 \\ \hline
{\bf Co$_{5}$/Au} & 0.24 & $1.22\cdot10^{-2}$ & 4.94 & $5.07\cdot10^{-2}$ & 20.52 \\ \hline
{\bf Co$_{10}$/Cu} & 0.59 & $5.33\cdot10^{-3}$ & 0.90 & $1.06\cdot10^{-2}$ & 1.81 \\ \hline
{\bf Co$_{10}$/Ag} & 0.63 & $4.37\cdot10^{-3}$ & 0.69 & $1.31\cdot10^{-2}$ & 2.06 \\ \hline
{\bf Co$_{10}$/Pd} & 0.33 & $8.89\cdot10^{-3}$ & 2.67 & $2.05\cdot10^{-2}$ & 6.14 \\ \hline
{\bf Co$_{10}$/Pt} & 0.37 & $5.02\cdot10^{-3}$ & 1.37 & $1.25\cdot10^{-2}$ & 3.41 \\ \hline
{\bf Co$_{10}$/Au} & 0.24 & $1.05\cdot10^{-2}$ & 4.42 & $3.69\cdot10^{-2}$ & 15.53 \\ \hline
{\bf Ni$_5$/Cu} & 0.69 & $3.11\cdot10^{-3}$ & 0.45 & $8.31\cdot10^{-3}$ & 1.21 \\ \hline
\end{tabular}
\caption{Conductance and Conductance Oscillations for different metallic multilayers: majority band}
\label{tavola4}
\end{table}
\end{center} 

\newpage

\begin{center}
\begin{table}[h]
\begin{tabular}{|c|c|c|c|c|c|} \hline
\ Multilayer \ & \ \ \ $\Gamma$  \ \ \ & \ \ \ $\Delta\Gamma$  \ \ \ &
 \ \ \  ${\Delta\Gamma}/{\Gamma}  $ (\%) \ \ \ & \ \ \ $\Delta\Gamma_{\rm max}$
 \ \ \ & \ \ \  ${\Delta\Gamma_{\rm max}}/{\Gamma}$ (\%) \ \ \ \\ \hline
{\bf Co$_{5}$/Cu} & 0.29 & $1.19\cdot10^{-2}$ & 3.97 & $4.21\cdot10^{-2}$ & 14.04 \\ \hline
{\bf Co$_{5}$/Ag} & 0.28 & $1.15\cdot10^{-2}$ & 4.08 & $4.01\cdot10^{-2}$ & 14.31 \\ \hline
{\bf Co$_{5}$/Pd} & 0.18 & $1.23\cdot10^{-2}$ & 6.93 & $4.13\cdot10^{-2}$ & 23.24 \\ \hline
{\bf Co$_{5}$/Pt} & 0.19 & $8.54\cdot10^{-3}$ & 4.29 & $2.51\cdot10^{-2}$ & 12.65  \\ \hline
{\bf Co$_{5}$/Au} & 0.20 & $1.06\cdot10^{-2}$ & 5.08 & $3.52\cdot10^{-2}$ & 16.86 \\ \hline
{\bf Co$_{10}$/Cu} & 0.32 & $1.08\cdot10^{-3}$ & 3.38 & $2.82\cdot10^{-2}$ & 8.80 \\ \hline
{\bf Co$_{10}$/Ag} & 0.32 & $1.75\cdot10^{-2}$ & 5.54 & $5.73\cdot10^{-2}$ & 18.15 \\ \hline
{\bf Co$_{10}$/Pd} & 0.16 & $1.56\cdot10^{-2}$ & 9.78 & $3.50\cdot10^{-2}$ & 21.14 \\ \hline
{\bf Co$_{10}$/Pt} & 0.19 & $9.02\cdot10^{-3}$ & 4.71 & $2.63\cdot10^{-2}$ & 13.73 \\ \hline
{\bf Co$_{10}$/Au} & 0.16 & $9.60\cdot10^{-3}$ & 5.95 & $2.34\cdot10^{-2}$ & 14.53 \\ \hline
{\bf Ni$_5$/Cu} & 0.51 & $7.63\cdot10^{-3}$ & 1.49 & $2.39\cdot10^{-2}$ & 4.71 \\ \hline
\end{tabular}
\caption{Conductance and Conductance Oscillations for different metallic multilayers: minority band}
\label{tavola5}
\end{table}
\end{center} 

\newpage

\begin{center}
\begin{table}[h]
\begin{tabular}{|c|c|c|c|c|c|} \hline
\ Multilayer \ & \ \ \ $\Gamma$  \ \ \ & \ \ \ $\Delta\Gamma$  \ \ \ &
 \ \ \  ${\Delta\Gamma}/{\Gamma}$ (\%) \ \ \ & \ \ \ $\Delta\Gamma_{\rm max}$
 \ \ \ & \ \ \  ${\Delta\Gamma_{\rm max}}/{\Gamma}$ (\%) \ \ \ \\ \hline
{\bf Co$_{5}$/Cu} & 0.16 & $5.33\cdot10^{-3}$ & 3.31 & $1.35\cdot10^{-2}$ & 8.40 \\ \hline
{\bf Co$_{5}$/Ag} & 0.18 & $6.71\cdot10^{-3}$ & 3.62 & $2.34\cdot10^{-2}$ & 12.68 \\ \hline
{\bf Co$_{5}$/Pd} & 0.13 & $7.98\cdot10^{-3}$ & 6.04 & $1.52\cdot10^{-2}$ & 15.00 \\ \hline
{\bf Co$_{5}$/Pt} & 0.14 & $6.87\cdot10^{-3}$ & 4.81 & $2.28\cdot10^{-2}$ & 16.01 \\ \hline
{\bf Co$_{5}$/Au} & 0.11 & $1.25\cdot10^{-2}$ & 10.85 & $5.49\cdot10^{-2}$ & 47.45 \\ \hline
{\bf Co$_{10}$/Cu} & 0.18 & $6.40\cdot10^{-3}$ & 3.51 & $1.62\cdot10^{-2}$ & 8.94 \\ \hline
{\bf Co$_{10}$/Ag} & 0.21 & $5.10\cdot10^{-3}$ & 2.46 & $1.06\cdot10^{-2}$ & 5.17 \\ \hline
{\bf Co$_{10}$/Pd} & $9.41\cdot10^{-2}$ & $1.09\cdot10^{-2}$ & 11.62 & $3.15\cdot10^{-2}$ & 33.45 \\ \hline
{\bf Co$_{10}$/Pt} & 0.10 & $4.77\cdot10^{-3}$ & 4.69 & $1.43\cdot10^{-2}$ & 14.12 \\ \hline
{\bf Co$_{10}$/Au} & $8.40\cdot10^{-2}$ & $6.95\cdot10^{-3}$ & 8.27 & $1.66\cdot10^{-2}$ & 19.77 \\ \hline
{\bf Ni$_5$/Cu} & 0.47 & $4.85\cdot10^{-3}$ & 1.02 & $1.38\cdot10^{-2}$ & 2.91 \\ \hline
\end{tabular}
\caption{Conductance and Conductance Oscillations for different metallic multilayers: AF configuration}
\label{tavola6}
\end{table}
\end{center}

\end{document}